\documentclass[amsfonts,amssymb,amsmath,twocolumn,superscriptaddress]{revtex4-1}

\usepackage{braket}
\usepackage{graphicx,color}

\def\({\left(}
\def\){\right)}

\begin{document}
\title{Quantum key distribution protocols with slow basis choice}
\author{Toshihiko Sasaki}
\affiliation{Photon Science Center, Graduate School of Engineering, The University of Tokyo, Bunkyo-ku, Tokyo 113-8656, Japan}
\author{Kiyoshi Tamaki}
\affiliation{NTT Basic Research Laboratories, NTT Corporation, 3-1, Morinosato Wakamiya Atsugi-Shi, Kanagawa, 243-0198, Japan}
\author{Masato Koashi}
\affiliation{Photon Science Center, Graduate School of Engineering, The University of Tokyo, Bunkyo-ku, Tokyo 113-8656, Japan}

\begin{abstract}
 Many quantum key distribution (QKD) protocols require  random choice of measurement basis for each pulse or each train of pulses.
 In some QKD protocols, such as  the Round-Robin Differential Phase Shift (RRDPS) QKD protocol, 
 this requirement is a bit challenging as randomly choosing hundreds of settings for every, say, 100 pulses may be too fast with
 current technologies.
 In this paper, we solve this issue by proving the security of QKD protocols with slow basis choice without compromising the secret key rate.
 We also show that the random choice of the bases for the state preparation can be made slow if the signals do not leak any information on the basis.
 Examples of QKD protocols that our technique  can apply include the RRDPS protocol and BB84-type protocols, and our technique relaxes demands
 for the implementation of QKD systems.
\end{abstract}

\maketitle
\section{Introduction}
Quantum key distribution (QKD) \cite{Bennett1984,Ekert1991,Bennett1992b,Brus1998,Mayers1998,Grosshans2002,Inoue2003,Scarani2004,Stucki2005,Sheridan2010, Lo2012} is an informationally secure way to share common random bits (the secret key) between the two parties, called Alice and Bob.
Since QKD is a physical cryptography, its security proofs need to assume some mathematical models of the devices used in QKD systems.
In particular, for simplicity of the proof, those models are sometimes a bit demanding to implement in practice, and this issue has to be solved for the actual implementation of the protocol.
For instance, a perfect singe-photon source was assumed in the security proof in \cite{Shor2000}, which is beyond current technologies, and this assumption was later removed by the so-called tagging idea,
and now we are allowed to use a coherent light source \cite{Gottesman2004}.

One of such demanding assumptions is that the measurement basis is chosen randomly  for each of the incoming pulses
in order to prevent an eavesdropper, Eve, from freely reading out the information.
For example, in the case of the Round-Robin Differential Phase
Shift (RRDPS) QKD protocol \cite{Sasaki2014,Takesue2015a}, which is a recently
proposed protocol, the measurement setting has to be randomly chosen
from hundreds of settings for each of train of pulses. Thanks to
this measurement scheme, this protocol has a distinguished feature
that the privacy amplification is totally independent of the signal
disturbance.
Unfortunately, however, its implementation, especially the measurement scheme with more than 100 settings,
is challenging in practice, although it is not impossible with elaborating
setups \cite{Guan2015,Wang2015,Li2015}.
Therefore, it would significantly make the implementation simpler
if we can get rid of this requirement of the fast optical switch from the experiment.

In this paper, we remove the necessity of the demanding fast switch by proving the security of the RRDPS protocol with slow basis choice,
and in the modified protocol we are allowed to switch the setting as frequent as the occurrence of the detection events, rather than the system clock rate.
The proof technique can apply for the BB84-type protocol as well, and we also show that the random choice of the bases for the state
preparation can be made slow if the signals do not leak any information on the basis.
We note that slow basis choice has an additional benefit of reducing the amount of the random numbers that is used for the setting choices,
which makes the requirements for QKD systems even less demanding.

This paper is organized as follows.
In Sec.\ \ref{sec-pot-vul}, we show that a naive slow basis choice of the measurement setting renders a BB84 protocol insecure by explicitly introduce an eavesdropping strategy.
In Sec.\ \ref{sec-bb84}, we introduce a secure BB84 protocol with slow basis choice, and show its security.
Our security proof is simple and applicable to other QKD protocols, including the RRDPS protocol. 
In Sec.\ \ref{sec-rrdps}, we take the RRDPS protocol with the photon-number-resolving detectors as an example to examine
how the slow basis choice affects the secret key rate.
In Sec.\ \ref{sec-con}, we summarize this paper.
In Appendix, we analyze the effect of the initialization process of the devices and the the threshold detectors having the dead time  in the case of the RRDPS protocol.

\section{Potential vulnerability}
\label{sec-pot-vul}
In this section, as an example that a naive slow choice of the measurement setting results in breakage of the security,
we consider a particular BB84 protocol with single-photon sources.
Before presenting the protocol, we introduce some terminologies and list up assumptions we make.
First, we call the two complementary bases in the BB84 protocol as the $Z$ basis and the $X$ basis, and we assume that Alice's source is a single-photon source and they use the $Z$ basis to generate the final key. 
Moreover, as is the case in many security proofs \cite{Lo1999, Shor2000, Renner2005, Koashi2009}, we assume that the
detection efficiency of Bob's measurement apparatus is independent of the measurement basis. 
With these assumptions, we describe the BB84 protocol with a naive slow choice of the measurement 
setting, {\it i.e.,} Bob chooses his measurement basis only for every $M$ time slots.

\begin{enumerate}
 \item    Alice randomly selects the basis from the $Z$ basis or the $X$ basis,
					and she also randomly chooses one of the eigenstates in the chosen basis to encode a bit in the single-photon pulse.
					Then, she sends the  pulse to Bob over a quantum channel.
					She repeats this process many times.
 \item    Bob randomly chooses the measurement basis from the $Z$ basis or the $X$ basis for each sequence of the incoming pulses, which contains $M$ pulses.
 \item    By using an authenticated classical public channel, Alice and Bob keep those instances where Bob detects the signals and 
            Alice and Bob have chosen the
            same bases. After this sifting step, they perform error correction and privacy amplification to generate the final key.
\end{enumerate}
In what follows, in order to see that the final key in this protocol is insecure, we will explicitly construct 
an eavesdropping strategy against this protocol. For this, we define $p_Z$ ($p_X$), $\eta, N_{\mathrm{tot}}, N, e_{Z},$ and $, e_{X}$ as the probability to choose
the $Z$ ($X$) basis, the probability for Bob to detect a photon sent from Alice, {\it i.e.,} the transmission rate,
the number of pulses sent by Alice, the length of the sifted key length, the error rate in the $Z$ basis, 
and the 
error rate in the $X$ basis, respectively. To be more specific, we set these parameters and the length $M$ of 
the sequence as $p_Z=p_X = 0.01, \eta = 10^{-6}, N_{\mathrm{tot}}= 10^{13}$, {and} $M=10^5$, respectively.
With these numbers, we may expect the length $N$ of the sifted key to be $10^{7}$
and Bob detects a signal out of $10^{6}$ pulses on average under the normal operation without Eve.
In order to let Eve possess the power to cause the detection event
at Bob's side whenever she pleases, we assume that all the losses are under Eve's control. Under these assumptions, 
we consider the following eavesdropping strategy. 
\begin{enumerate}
 \item Eve randomly chooses 99 sequences and 1 sequence out of $10^8$ sequences, temporarily stores all the pulses in the chosen 100 sequences in her lab, and 
then blocks and discards all the rest of the pulses.
 \item Eve performs measurements to all the pulses in the 99 sequences with the $Z$ basis, prepares 
 pulses in the states corresponding to the measurement outcomes, and then sends all of the prepared pulses to Bob.
 \item Eve sends all the pulses in the 1 sequence to Bob without performing any operation to them.
\end{enumerate}
Now, we show that this eavesdropping renders the final key insecure given that Bob does not monitor 
the bunching of the detection events. For this, first note that the transmission rate is preserved as Eve sends $100M(=10^7)$ pulses to Bob in total.
Next, we calculate the probability that Alice and Bob end up with an identical final key without noticing this eavesdropping, {\it i.e.,}
the joint probability that Bob chooses the $Z$ basis for the sequences that Eve has measured with the $Z$ basis 
and he chooses the $X$ basis for the  sequence without Eve's disturbance. It is straightforward to obtain this probability, and this is given by 
$0.99^{99}\times 0.01 \sim 0.0004$.
Note that this probability is higher than the typical value of the security parameter \cite{Mueller-Quade2009}, like $10^{-10}$.
Therefore, we conclude that the final key fails to 
meet the typical security criteria,
implying the breakage of the security.

It is clear that this eavesdropping works out because Alice and Bob ignore the bunching of the detection events,
and this attack should be detected by monitoring this bunching.
In the next section, we will make this intuition more rigorous and present the explicit modifications that we need to make the protocol secure .

\section{Security proof for BB84 protocol with slow basis choice}
\label{sec-bb84}
In this section, we modify  the protocol presented in the previous section and
show the security of the new protocol. The modification is made 
between Step 2 and Step 3 in the previous protocol, which is:
\begin{enumerate}
 \item[2.5] If Bob obtains more than 1 detection events in a sequence, then he discard the sequence.
\end{enumerate}
To see that  the protocol with this additional step is secure, we consider the  protocol from
the viewpoint of the so-called entanglement based protocol \cite{Lo1999, Shor2000}.
The entanglement based protocol we will introduce is equivalent to the modified protocol in the sense that all the quantum and classical information accessible to Eve are exactly the same, and Alice and Bob's classical data remains the same.
Therefore, Eve cannot behave
differently between the two protocols, and we can use the entanglement based protocol for the security proof. In particular,
we consider the following entanglement based protocol.
\begin{enumerate}
 \item     Alice prepares a qubit and makes it entangled with the single photon in such a way that the photon randomly becomes one of
             the $Z$ ($X$) basis eigenstates if
			 the qubit is measured with the $Z$ ($X$) basis.
			 She sends the photon to Bob over the quantum channel while she keeps the qubit on her side. She repeats this process many times.
 \item    Bob measures the number of photons in each pulse,
					counts the number of pulses that contains non-zero photon in each of the sequences, and broadcasts the latter numbers for each sequence over the authenticate classical channel.
 \item  Alice and Bob keep those sequences where  Bob detects photons only in one pulse, and they discard all the other sequences. 
 \item  By using the authenticated classical channel, Bob tells Alice the position of the pulse with the photons for each of the surviving sequence.
 \item According to the information from Bob, Alice discards all her qubits except those corresponding to the pulses containing the photons in one pulse.
 \item For each pair of the qubit and the pulse, Alice and Bob randomly choose their bases and measure their systems with the chosen bases.
 \item Alice and Bob announce the basis they have used for each pulse over the authenticated classical channel.
 \item   With the help of the authenticated classical public channel, Alice and Bob keep those instances where they have chosen the
				 same basis. After this sifting, they perform the estimation of the bit error rate as well as the leaked information, followed by error correction and privacy amplification to obtain the final key.
\end{enumerate}

The important point in this protocol is that Alice and Bob's bases choice are independently made
{\it after} the post-selection of the qubits and pulses in Step 6. This delayed choice of the
measurement bases is the key in the security proof, and this idea is, in fact, used in many security proofs for 
BB84 protocol \cite{Lo1999, Shor2000, Renner2005, Koashi2009}. For instance, in the Shor-Preskill's security proof \cite{Shor2000},
the intuition of the proof is that this delayed choice enables Alice and Bob to symmetrize the shared quantum states without 
being affected by Eve's strategy, and this symmetry is enough to prove the security.
Moreover, the security proof based on the complementarity scenario or the entropic uncertainty relationship \cite{Koashi2009,Renner2005} enables us to get rid of the qubit
assumption at Bob's side, which is made in the Shor-Preskill's proof, allowing us to use threshold detectors.
Therefore, by directly applying such security proofs, one can see that the above protocol is secure.

The above protocol can be simplified furthermore.
First, Alice can use a fixed basis over pulses in a sequence
given that the single-photon source does not leak any information on the basis.
This is so because the security of the above protocol relies also on the fact that Eve cannot obtain any information 
on Alice's bases prior to her attack.
As long as this condition is satisfied, the protocol is secure.

Next, observe that the fixed length of the sequence $M$ does not play any role in the security proof when we use the single photon source.
The only important thing is that each post-selected sequence contains only one detection event. This leads us to the following further modification of the
protocol: Bob immediately makes the announcement upon obtaining the detection event, and then Alice and Bob randomly
choose the bases just after this announcement.
This modification is secure because we still have only one detection event for
each sequence, whose length is now depending on the occurrence of the detection events.
This way we can reduce the
number of the sequences discarded in the protocol with a single-photon source and we can improve the secret key generation rate without compromising the security.

So far, we have assumed the single-photon source, however,
the practical implementations often employ a coherent light source.
Note that we can apply our analysis to this practical case if we employ the
the decoy state method \cite{Hwang2003,Lo2005,Wang2005}.
This is so because this method allows us to estimate the fraction of the detection events
caused by the sequences where Alice has emitted only a single-photon,
and this estimation is crucial for our proof to apply.

To summarize, Bob can choose
his measurement basis  after the detection event, and if the source does not leak any information on the
basis, then Alice is also allowed to choose her basis just after Bob's detection event. 
We note that since this argument does not rely on what basis set we use, we can also prove 
the security of other protocols, such as the loss-tolerant protocols with three states \cite{Tamaki2014}, 
with slow basis choice. In the case of the loss-tolerant protocol with three states, however, Bob's choice can be made slow 
but Alice's choice cannot be made slow as her sending state leaks some basis information to Eve. 
In the next section, we apply our idea to another protocol, the RRDPS protocol.

\section{secret key rate of RRDPS protocol with slow basis choice}
\label{sec-rrdps}
In this section, we first apply our idea to a RRDPS protocol with slow basis choice, and then we examine how the slow choice of the measurement basis affects the secret
key generation rate of the RRDPS protocol. 
Before we describe the RRDPS protocol, we define some terminologies. We call bunches of pulses a block, and we name
collections of blocks as a sequence. Therefore, the block corresponds to the pulse in the BB84 in the previous section.
Keeping these terminologies in our mind, the RRDPS protocol with the slow choice of the measurement basis runs as follows.
\begin{enumerate}
 \item    Alice emits a train of weak coherent laser pulses with an interval $T$ to form a block consisting of $L$ pulses,     
			 applies a phase shift randomly chosen from $\{0,\pi\}$ on each pulse, and then sends 
             the block to Bob through the quantum channel. Alice repeats this many times.
 \item    For each sequence that consists of $M$ blocks, 
            Bob first randomly chooses a number corresponding to a delay time out of 
             $\{T, 2T, \cdots, (L-1)T\}$. Then, he employs the chosen number as the delay time in a variable-delay interferometer to interfere 
             the pulses in $M$ blocks, {\it i.e.,} in the sequence, and uses the photon-number-resolving detectors to detect signals. 
 \item    For each sequence, Bob keeps only one block if it is the first block with the detection event and the block contains only one photon.
					   Otherwise, he discards the sequence. Bob announces to Alice over the authenticated classical channel which sequences have been discarded.
 \item    For each sifted sequence, Bob records the bit value depending on the measurement outcome and announces over the authenticated classical channel
            the delay time used in Step 2 as well as the position of the surviving block in the sequence.
 \item   By using the information from Bob, Alice computes the bit values for each of the sifted sequence.
 \item   Alice and Bob perform error correction and privacy amplification to obtain the final key.
\end{enumerate}
Note in this protocol that $M$ represents how slow the choice of the measurement basis is, and $M=1$ corresponds to the
case of the original RRDPS protocol.

The security of this  RRDPS protocol with slow basis choice can be justified with the same manner as
the one of the BB84-type protocols in the previous section.
That is, the measurement basis choice for the surviving block for the detection event can in principle be delayed {\it after}
the detection of the photons by a fictitious QND measurement \cite{Takesue2015a} measuring the photon number of the block,
and this basis choice is random as is the case for the original RRDPS protocol.
It means that we can use the same security proof, and the similar key rate formula to the original protocol can be achieved \cite{Sasaki2014}.
Having said this, however, note that in the original proof, the blocks are evaluated by $e_{\text{src}}$, which is the rate of the blocks which contains more than $\nu_{\text{th}}$ photons at Alice's side.
In our protocol, we have to replace $e_{\text{src}}$ with $e_{\text{src,slow}}$,
which is the probability that there is at least one block containing larger than $\nu_{\mathrm{th}}$ photons in a sequence at Alice's side,
and it is written as
\begin{equation}
 \label{esrc_esrcslow}
 e_{\text{src,slow}} = 1 - (1-e_{\text{src}})^M.
\end{equation}
By directly borrowing the security proof of the original RRDPS protocol, we have the secret key generation rate $G$ per pulse in our RRDPS protocol as
\begin{align}
 \label{keyrate}
	 G &= \frac{Q}{ML}
 \( 1- h(e_{\mathrm{bit}})   - h\( e_{\mathrm{ph}}\) \),\\
  \label{phase}
 e_{\mathrm{ph}} &= \frac{e_{\mathrm{src,slow}}}{Q} + \(1-\frac{e_{\mathrm{src,slow}}}{Q}\)\frac{\nu_{\mathrm{th}}}{L-1},
\end{align}
where $Q$ is the detection rate per sequence, $e_{\mathrm{bit}}$ is the bit error rate,
and $e_{\mathrm{ph}}$ is  the so-called phase error rate that is used in privacy amplification to generate the secret key \cite{Shor2000,Koashi2009,Sasaki2014}.

To simulate the resulting secret key generation rate of our RRDPS protocol with the block size $L=128$, we assume the following channel model.
The experimental setup has an inevitable system error, whose error rate is 
$e_{\mathrm{sys}}$, and this error is independent of the channel transmission and the dark count rate $d_c$ of the detectors,
and $e_{\mathrm{sys}}$ and $d_{\mathrm{c}}$ are assumed to be $0.03$ and $10^{-9}$, respectively.
As for Alice, she  uses a coherent light source, and it leads to the expression of $e_{\text{src}}$  as $1-e^{-L\mu}\sum_{\nu=0}^{\nu_{\mathrm{th}}}(L\mu)^\nu/\nu!$,
where $\mu$ is the average photon number per pulse.
The explicit expressions of $Q$ and $e_{\text{bit}}$ under these assumptions are provided in Appendix.
Under these assumptions, in Fig.\;\ref{keyrate-graph}, we show the resulting secret key generation rate per pulse as a function of
the transmission rate of the quantum channel between Alice and Bob,
where we vary the size $M$ of the sequence from $1$ to $10^6$.
We have optimized  $\mu$ and  $\nu_{\mathrm{th}}$ in the simulations.
Note that the above protocol assumes the use of photon-number-resolving 
detectors, however, we show in Appendix that we can use the threshold detectors having the dead time without
frustrating the key rate (see Appendix and Fig.\;\ref{keyrate-graph2}).
Moreover, we can accommodate the effect of the initialization process of the devices and its effect is shown not to be very small (see Fig.\;\ref{keyrate-graph3}).

We find in Fig.\;\ref{keyrate-graph} that the key rates for each $M$ becomes constant when the channel transmission rate $\eta$ is high.
On the other hand, when $\eta$ is low, all the key rates converge to the one of $M=1$.
This clearly shows that we can safely adopt the slow basis choice without making any compromise  in the high loss regime.
In order to explain these tendencies in the normal operation without eavesdropping, we consider two extreme cases $M\eta \gg 1$ and $M\eta \ll 1$.
Here, $M\eta$ is regarded as the effective transmission rate per block because the mean photon number of a sequence after the transmission is given by $M\eta L\mu$.
In the case of $M\eta\gg 1$, that is, in the high transmission regime, the detection rate is saturated, and therefore the key rate is
also saturated because the detection rate is the key factor determining the key generation rate, which intuitively explains the saturation in Fig.\;\ref{keyrate-graph}.
On the other hand, in the case of $M\eta\ll 1$, the mean photon number of each of the pulse is so small that the multi-detection in the same sequence rarely happens.
It means that the detection rate does not depend on the size of the sequence or the block,
and it explains the convergence in Fig.\;\ref{keyrate-graph}.
We remark that since the above discussion does not explicitly use the properties of the RRDPS protocol, these tendencies should hold also in other QKD protocols.

\begin{figure}
	\begin{center}
	 \includegraphics[width=7cm]{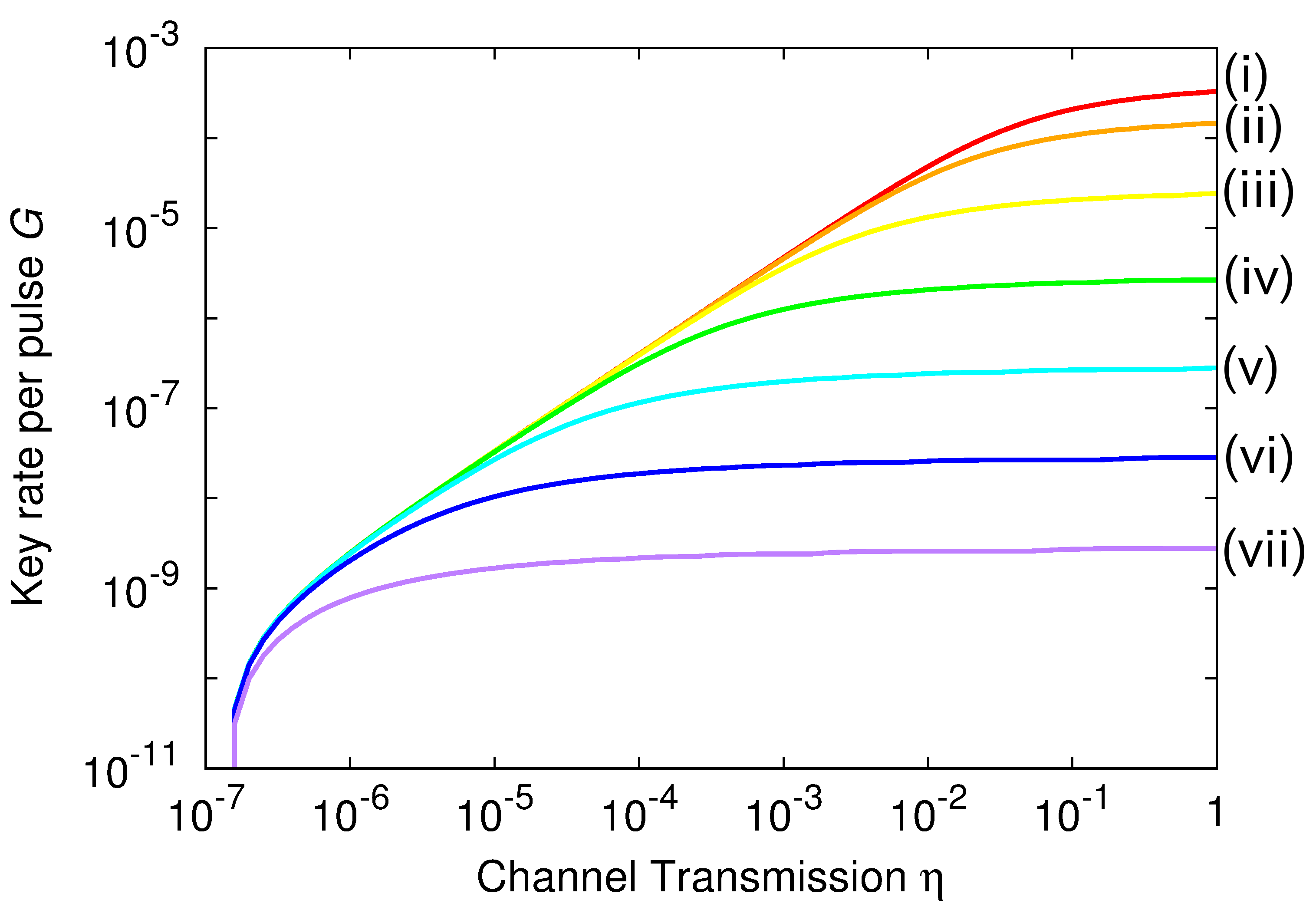}
	\caption{The secret key generation rate of the RRDPS protocol with slow basis choice and photon-number-resolving detectors versus channel transmission.
	Lines labeled {by} (i)-(vii) represent {the case of the sequence with the length of} $1,10, 10^2,10^3,10^4,10^5,$ and $10^6$, respectively.}
\label{keyrate-graph}	
	\end{center}
\end{figure}

\section{Summary and discussion}
\label{sec-con}
In this paper, we first show that a naive slow basis choice for the BB84 protocol results in the generation of an insecure key by explicitly
constructing an eavesdropping. Then, we see that this vulnerability can be circumvented by a slight modification
of the sifting process in the post-processing of the protocol, and we prove the security of the modified protocol.
Our proof does not exploit the detailed structure of the protocol, and it can apply to other protocols, such as the loss-tolerant protocols.
We also show that random choice of bases for the state preparation can be made slow for those protocols if
the signals does not leak any information on the basis.
Finally, in order to show how the slow basis choice affects the secret key generation rate, we simulate the resulting secret key generation rate 
of the RRDPS protocol with slow basis choice.
It follows from the simulation results and the key formula that the choice can be made as slow as Bob's detection rate of the pulse without compromising the key rate in typical experiments.

In this paper, we do not consider Bob who processes the data from a bunch of pulses that contains more than one photon,
but instead Bob who discards such data. We may be able to accommodate such data in the security proof by modifying the
post-processing furthermore. Having said that, however, the accommodation of such data does not significantly increase the secret key generation rate so much
as the number of such data is very low compared to the one of the data from only one photon. Moreover,
it is doubtful if it is still beneficial to consider such a rare event in practice cases where finite-size effects have to be accommodated.
In this case, we usually need to send huge number of pulses to obtain a positive gain, which is not practical, and therefore,
we do not take into account such data in the present work.

\begin{acknowledgements}
We thank H. Takesue, K. Azuma, W. Munro, G. Knee, and F. Furrer for valuable discussions.
This work was funded in part by ImPACT Program of Council for Science, Technology and Innovation (Cabinet Office, Government of Japan),  
Photon Frontier Network Program (MEXT).
\end{acknowledgements}

\appendix*

 \section{}

 In this appendix, we consider
 the effect of the initialization process of the slow device and how to use the threshold detectors with the dead time.
We also provide the explicit expressions of the parameters needed for the simulations.
 
 First, we consider the initialization process.
 In the actual experiment, we sometimes need to wait a certain period of time after every sequence.
 For example, if we use the slow optical switches to change the basis,  we need to wait
 until the switching process is completed.
 We also have to consider the effect of the dead time of the photon detectors.
 All devices have to be initialized before the next sequence, and we discard
the data obtained during the initialization process because the assumptions made in the security proof are not satisfied during the process.
 Now suppose that all devices are initialized
in a fixed period of time from the end of each sequence,
 and let $c_d$ be the number of pulses emitted by the light source during the  period.
Note that  we have to be careful about the initialization of the photon detectors in a fixed period because the dead time can be prolonged by the additional input \cite{Lydersen2011a, Fujiwara2013}.
In order to avoid this problem, we assume that all inputs into the photon detectors are blocked during the initialization process.
 Since all the pulses during the initialization process are blocked, the security proof for the key from the sequence can apply without any modification.
 The only thing we have to change is  the factor $ML$ in Eq.\;(\ref{keyrate})
 because Eq.\;(\ref{keyrate}) represents the key rate {\it per pulse},
and the effect of the initialization must be taken into account in this rate. 
 Thus, we replace the factor $ML$ in Eq.\;(\ref{keyrate}) with $ML+c_d$,
 but the other factors, such as $M$ in  Eq.\;(\ref{esrc_esrcslow}), do not change.
 
 \begin{figure}
		\begin{center}
		 \includegraphics[width=7cm]{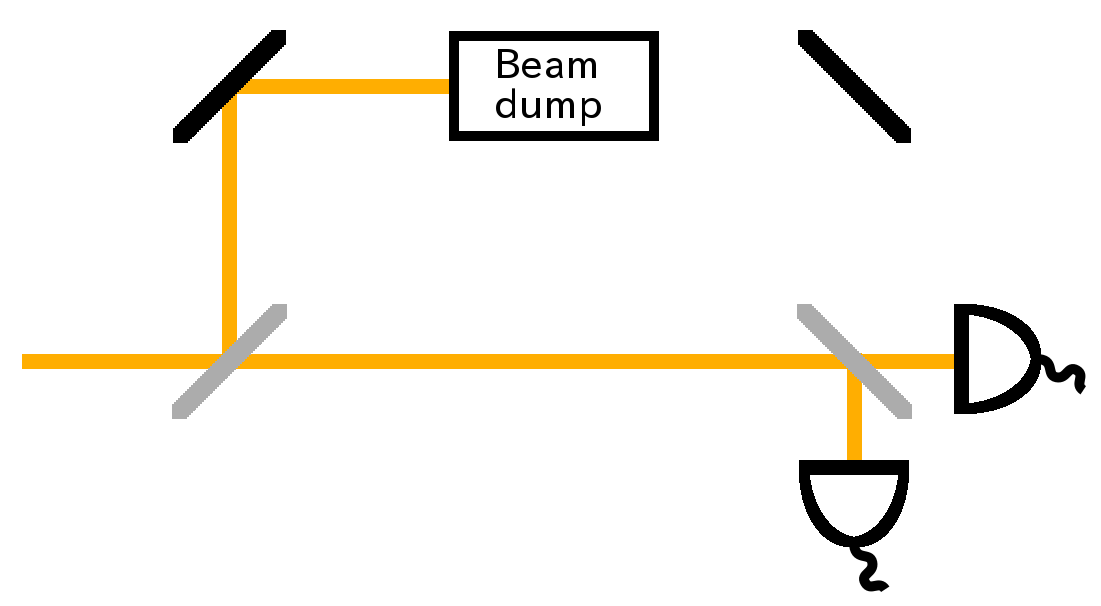}
	 \caption{A measurement setup to estimate the number of sequences that include more than one photon.
		 The beam dump simply blocks all the pulses along the longer path that is used in the
                 standard measurement. }
	 \label{fig-drop-delay-path}
		\end{center}
 \end{figure}

\begin{figure}
	\begin{center}
	 \includegraphics[width=7cm]{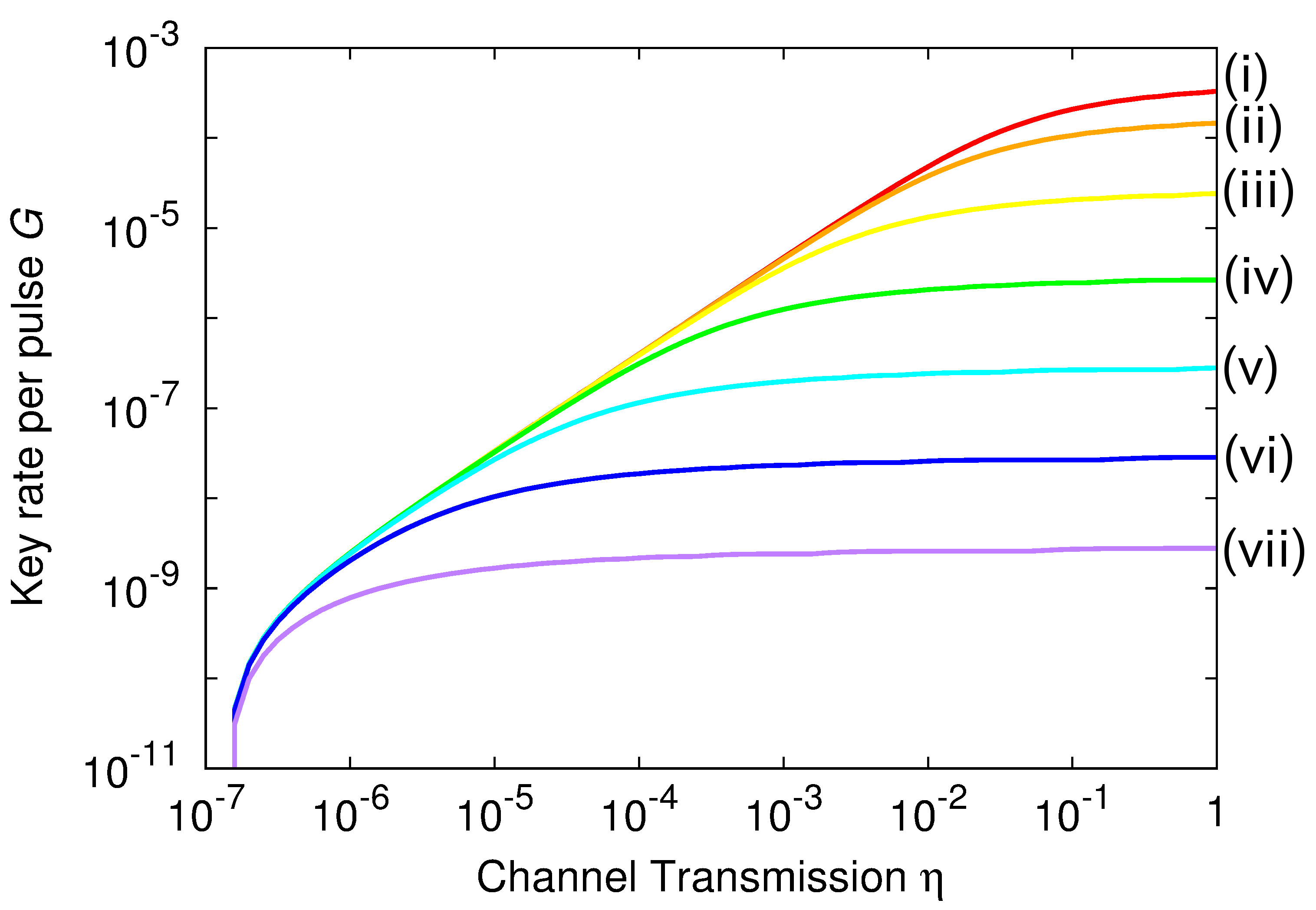}
	\caption{The secret key generation rate of the RRDPS protocol with slow basis choice and the threshold  detectors  versus channel transmission.
	Lines labeled {by} (i)-(vii) represent {the case of the sequence with the length of} $1,10, 10^2,10^3,10^4,10^5,$ and $10^6$, respectively.}
\label{keyrate-graph2}	
	\end{center}
\end{figure}
\begin{figure}
		 \begin{center}
			\includegraphics[width=7cm]{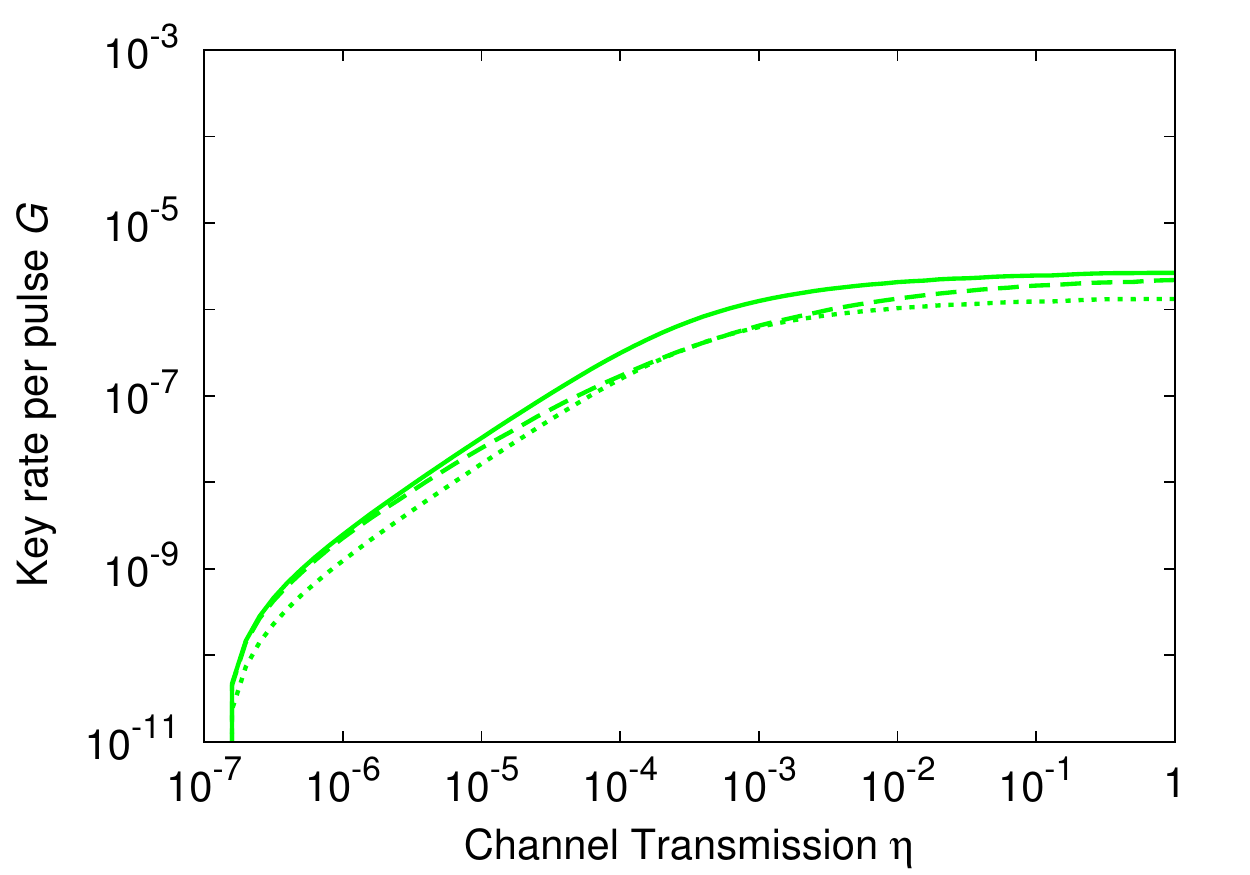}
	\caption{The secret key generation rate of the RRDPS protocol with slow basis choice and the threshold  detectors having the dead time versus channel transmission.
		A solid line, a dashed line, and a dotted line represent the rate for  $c_d=0$ and $M=10^3$, for $c_d=1.28\times 10^5$ and the optimal $M$ for each $\eta$, and for $c_d=1.28\times 10^5$ and $M=10^3$, respectively.}
\label{keyrate-graph3}	
		 \end{center}
\end{figure}

Next, we consider the use of the threshold detectors
rather than the photon-number-resolving detectors assumed in the original RRDPS protocol as well as the one in the main text.
To prove the security in this scenario, first note that
it loses no generalities if we assume Bob who performs a fictitious QND measurement \cite{braginskiui1975quantum,Unruh1979,Braginsky1980,Caves1980} counting the number of photons in each block. This is
so because the interferometer is a linear-optical device, and therefore, the QND measurement does not disturb Bob's measurement outcomes.
Therefore, we are allowed to regard the state of each block as a classical mixture of the Fock diagonal state, and our task is to estimate the
lower bound of the number of the sifted blocks containing only a single photon (note that the security proof of the original RRDPS protocol highly relies on the detection of a single photon).
As we will see below, this number is estimated from double-count events 
in a measurement scheme depicted in Fig.\;\ref{fig-drop-delay-path}, which is randomly chosen by Bob, and the beam dump blocks one of the paths in the interferometer.
As for the multiple photons in the sifted blocks, we employ the tagging idea in GLLP analysis \cite{Gottesman2004}, {\it i.e.,} we assume the worst 
case scenario that Eve has the maximum information on those instances.

For the estimation of the lower bound, Bob randomly chooses sequences and measures them with the measurement setup in Fig.\;\ref{fig-drop-delay-path}
instead of the variable-delay interferometer.
In this setup, we define the double-count event as follows. In this event, both of the detectors click within a block, and moreover, this event has to be the first among blocks with click events of at least one of the detectors.
If the dead time of the detectors were short enough to be negligible, we would also use multi-click events of the same detector within a block for the estimation.
Unfortunately, however, the dead time of  the actual photon detectors is not negligible, and we cannot necessarily use the second detection of the same detector.
Therefore, we conservatively adopt the above definition with the accommodation of only the first clicks of the two detectors in a block.
We note that even if the dead time is prolonged with a side-channel, it does not affect the rate of having the double-count given that the initialization process is properly completed after every sequence.

Now, we consider the relation between the double-count event and the sequence including more than one photon.
The probability that a photon, which is incoming to the input port of the measurement scheme in  Fig.\;\ref{fig-drop-delay-path},
is detected in either detector is 1/2.
When one of the detectors is in dead time, the probability that a photon is detected in the other detector is 1/4.
Thus, when there are more than one photon in a block, the probability of having the double-count event in the block is at least 1/8.
This means that the number of blocks including more than one photon can be lower bounded by eight times of the one of the double counts.

From these results, we can obtain the modified key rate formula as
 \begin{align}
	\label{newkeyrate}
	 G &= \frac{Q}{ML+c_d}
 \( 1- h(e_{\mathrm{bit}})  - \frac{e_{\text{mB}}}{Q}\right.\nonumber\\
	& \hspace{3cm} \left. - \(1-\frac{e_{\text{mB}}}{Q}\) h\( e_{\mathrm{ph}}\) \),\\
	\label{newphase}
 e_{\mathrm{ph}} &= \frac{e_{\mathrm{src,slow}}}{Q- e_{\text{mB}} } + \(1-\frac{e_{\mathrm{src,slow}}}{Q-e_{\text{mB}}}\)\frac{\nu_{\mathrm{th}}}{L-1},
 \end{align}
where $e_{\mathrm{mB}}$ is eight times the rate of the double-count events at Bob's side per sequence.

Next, we provide the explicit expressions of $Q, e_{\text{bit}}$, and $e_{\text{mB}}$
with the use of the threshold detectors and the accommodation of the initialization process
under the normal operation without Eve.
The main contribution of $Q$ is  a block containing one photon or a dark count after blocks including zero photon without dark counts.
Therefore, we have that
\begin{equation}
 Q \sim \sum_{m=0}^{M-1}\(e^{-L\eta\mu}(1-d_c)^{2L}\)^m \(\frac{1}{2}L\eta\mu e^{-L\eta\mu}+Ld_c\),
\end{equation}
where $\eta$ is a transmission rate, $\mu$ is an average photon number per coherent pulse and the factor $\frac{1}{2}$ corresponds to
the efficiency of delay line interferometer.
The bit error occurs when the system error or the dark count at the wrong output port happens.
This leads to the expression of $e_{\mathrm{bit}}$ as
\begin{equation}
\begin{split}
	&e_{\mathrm{bit}} \sim\\
	&\frac{\sum_{m=0}^{M-1}\(e^{-L\eta\mu}(1-d_c)^{2L}\)^m \(\frac{1}{2}L\eta\mu e^{-L\eta\mu}e_{\mathrm{sys}}+Ld_c\frac{1}{2}\)}{Q}.\\
 \end{split}
\end{equation}
Since both signal photons and dark counts can cause double counts,  $e_{\mathrm{mB}}$ is approximated as
\begin{equation}
 \begin{split}
	e_{\mathrm{mB}} \sim & 8\sum_{m=0}^{M-1}\(e^{-L\eta\mu}(1-d_c)^{2L}\)^m \(\frac{1}{8}\frac{1}{2}\(L\eta\mu\)^2 e^{-L\eta\mu} \right.\\
	&+\left.\frac{1}{2}L\eta\mu e^{-L\eta\mu}(2L-1)d_c + \frac{2L(2L-1)}{2}d_c^2\).
 \end{split}
\end{equation}

Now let us consider the effect of replacing the key rate formulae Eq.\;(\ref{keyrate}) and Eq.\;(\ref{phase}) with Eq.\;(\ref{newkeyrate}) and Eq.\;(\ref{newphase}).
The graph of the key rate formula Eq.\;(\ref{newkeyrate}) with $c_d=0$ is illustrated in Fig.\;\ref{keyrate-graph2}.
We can find that this graph is almost the same as the one presented in Fig.\;\ref{keyrate-graph}.
This is so mainly because $e_{\mathrm{mB}}$ is very small and negligible, and this means in particular that we can safely use the threshold detectors without frustrating the key rate.

Next, we consider the effect of a non-zero dead time $c_d$.
When $c_d$ is zero, the key rate always becomes larger by taking a smaller $M$.
When $c_d$ is not zero, however, it is not the case.
For instance, the key rate becomes larger by taking a larger $M$
if $M$ is not large enough to cause the saturation of the detection rate.
Therefore, we have to conduct the optimization of $M$ as well as $\mu$ and $\nu_{\mathrm{th}}$ for each $\eta$.
In Fig.\;\ref{keyrate-graph3}, we illustrate the key rate for $c_d=1.28\times 10^5$  with the optimal $\mu,\nu_{\text{th}},$ and $M$  as the dashed line when the other settings are the same as that in Fig.\;\ref{keyrate-graph2}.
We also plot the key rate for  $c_d=0$ and $M=10^3$ (solid line) and the one for $c_d=1.28\times 10^5$ and $M=10^3$ (dotted line), respectively.
As can be seen from Fig.\;\ref{keyrate-graph3}, we can obtain a nearly optimal key rate by setting $M$ to a fixed value satisfying $ML = c_d$.
This means that we set the time spent by a sequence to be equal to the dead time
although this simplification can cause the key rate to decrease from the optimal one by a factor of 1/2.

\end{document}